\documentstyle[12pt]{article}
\input epsf.sty
\def\ZZZ{{\hbox{ Z\kern-1.6mm Z}}}

\newcommand{\eps}{\epsilon}
\newcommand{\ra}{\rangle}
\newcommand{\la}{\langle}

\newcommand{\tl}{\lambda}

\newcommand{\sss}{{\cal L}_{av}}

\newcommand{\VV}{{\cal V}}

\newcommand{\AAA}{{\cal A}}

\newcommand{\OO}{{\cal O}}

\newcommand{\LL}{{\cal L}}

\newcommand{\square}{\Box}

\newcommand{\wt}{\widetilde}

\newcommand{\TT}{{\cal T}}

\newcommand{\be}{\begin{equation}}
\newcommand{\ee}{\end{equation}}
\newcommand{\ben}{\begin{eqnarray}\displaystyle}
\newcommand{\een}{\end{eqnarray}}
\newcommand{\refb}[1]{(\ref{#1})}
\newcommand{\p}{\partial}
\newcommand{\sectiono}[1]{\section{#1}\setcounter{equation}{0}}

\def\one{{\hbox{ 1\kern-.8mm l}}}
\def\zero{{\hbox{ 0\kern-1.5mm 0}}}
 
\begin{document}
{}~
{}~
\hfill\vbox{\hbox{hep-th/0403200}
}\break
 
\vskip .6cm
\begin{center}
{\Large \bf 
Energy Momentum Tensor and Marginal 

\medskip

Deformations 
in Open String Field 
Theory
}

\end{center}

\vskip .6cm
\medskip

\vspace*{4.0ex}
 
\centerline{\large \rm
Ashoke Sen}
 
\vspace*{4.0ex}

\centerline{\large \it Harish-Chandra Research Institute}

\centerline{\large \it  Chhatnag Road, Jhusi,
Allahabad 211019, INDIA}
 
\centerline{E-mail: ashoke.sen@cern.ch,
sen@mri.ernet.in}
 
\vspace*{5.0ex}
 
\centerline{\bf Abstract} \bigskip

Marginal boundary deformations in a two dimensional conformal field theory 
correspond to a family of classical solutions of 
the equations of motion of open string field theory. In this paper we 
develop a systematic method for relating the parameter labelling the 
marginal boundary deformation in the conformal field theory to the 
parameter labelling the classical solution in open string field theory. 
This is done by first constructing the energy-momentum tensor associated 
with the classical solution in open string field theory using Noether 
method, and then comparing this to the answer obtained in the conformal 
field theory by analysing the boundary state. We also use this method to 
demonstrate that in open string field theory the tachyon lump 
solution on a circle of radius larger than one has vanishing pressure 
along the circle direction, 
as is expected for a codimension one D-brane.

\vfill \eject
 
\baselineskip=18pt

\tableofcontents

\sectiono{Introduction and Summary} \label{sintro}

Methods of boundary conformal field theory have been used extensively to 
construct classical solutions in open string field 
theory. This exploits the correspondence between solutions of the 
equations of motion of open string field theory and conformally invariant 
boundary interactions in two dimenional
conformal field theories. One 
particularly important class of solutions obtained this way are those 
associated with 
marginal boundary deformations of the conformal field  
theory\cite{9402113,9404008,9811237,0203211,0203265}. Such deformations 
typically 
generate a continuous family of conformal field theories, and hence one 
expects that they will generate a continuous family of solutions of the 
open string field equations.

An algorithm for constructing 
such a family of solutions in open string field theory was 
developed in \cite{0007153,0107046,0304163} using the level 
truncation 
method\cite{KS,KP,9912249,0002237}. This involves choosing an arbitrary 
value of the open string field associated with the marginal operator, 
solving the equations of motion of all other fields, and substituting them 
back into the action. This generates an effective potential for the first 
field. It was demonstrated numerically that as we increase the level of
approximation, the effective potential becomes flatter. This suggests 
that in the full theory this effective potential is exactly flat, and 
hence we have 
a one parameter family of solutions labelled by the value of the open 
string field associated with the marginal operator.

This algorithm gives a way to generate a one parameter family of solutions 
in open string field theory if the corresponding conformal field theory 
has a marginal operator. However one of the problems which was not 
resolved in \cite{0007153} was to give an algorithm to systematically 
determine the relation between the marginal deformation parameter in the 
conformal field theory and the parameter labelling the solutions in open 
string field theory.
This is the problem we address in this paper. We work in the context of 
the 
specific class of marginal deformations analyzed in ref.\cite{0007153}, 
namely deformation by the first momentum mode of the tachyon on a 
D-brane in bosonic string theory wrapped on a circle of unit radius. Our 
main tool is to
compare
the energy-momentum tensor of the system in the two descriptions. In the 
conformal 
field theory description, the information about the energy-momentum tensor 
is contained in the boundary state\cite{9707068}. For the case at hand 
the exact 
expression for the boundary state of the system is known as a function of 
the marginal deformation parameter\cite{9402113,9811237} and using 
this 
one can compute the energy-momentum tensor of the system. In particular 
the pressure along the compact direction, given by the diagonal component 
of the stress tensor along this direction, has a simple expression in 
terms of the deformation 
parameter. We compare this with the corresponding result for the solution 
of the open string field equations. Matching the two results
determine the relationship between the parameter labelling the solutions 
of open string field theory and that labelling the boundary 
conformal field theories.

In order to implement this algorithm we need to construct 
the energy momentum tensor $T^\mu_\rho$ of open 
string field theory\cite{WITTENSFT} which contains non-local interactions. 
Energy-momentum 
tensor of non-local field theories was previously analyzed 
in 
\cite{0207107,0209197,0006235,0311184}, but we derive a different form of 
$T^\mu_\rho$ 
which is suitable for our analysis. We test our final expression by 
applying it 
to the usual scalar field theory, as well as to the $p$-adic string theory 
and reproducing the known answers up to the usual ambiguity of defining 
energy-momentum tensor using Noether prescription. We then apply it to 
open string field theory to compute the appropriate component of the 
$T^\mu_\rho$ associated with the family of solutions found in 
\cite{0007153}, and compare the results with the conformal field 
theory 
results to derive the relation between the conformal field 
theory parameter and string field theory parameter.

The paper is organised as follows. In section \ref{senergy} we
derive a form of the energy-momentum tensor for a general
translationally invarint non-local field theory. In section \ref{spadic} 
we use this to compute the energy of a time dependent classical solution 
in $p$-adic string theory and reproduce the answer given in 
\cite{0207107}. In section \ref{sappli} we use the expression for 
$T^\mu_\nu$ derived in section \ref{senergy} to compute the pressure 
associated with the lump solution in open string field theory on a 
D-string wrapped on a circle of  radius $>1$, and show that 
it
vanishes. This agress with the identification of the lump solution as the 
D0-brane located at a point on the circle, which is known to have 
vanishing pressure along the circle direction. This provides another 
consistency check for our formula. Finally in section \ref{smarginal} we 
apply our result to calculate the pressure associated with the family of 
lump
solutions in open string field theory wrapped on a circle of unit radius. 
Comparing this with the conformal field theory results, we then 
numerically 
determine the relation between the parameter labelling 
this family of solutions and the parameter labelling the corresponding 
conformal field theories. 

Although in this paper we 
shall be working in the specific context of bosonic string theory, the 
methods developed here are equally applicable to the analysis of classical 
solutions in open superstring field theory on unstable D-$p$-branes.

\sectiono{Energy Momentum Tensor of Non-local Field Theory} 
\label{senergy}

We begin by reviewing the general procedure for constructing a conserved 
current associated with a continuous symmetry. Let us consider a theory 
with scalar fields $\phi_1$, $\phi_2$, $\ldots$ in $p+1$ dimensions, and 
let $S[\{\phi_r\}]$ 
be a functional of these fields describing the action. Now suppose that 
$S$ is invariant under an infinitesimal symmetry transformation of the 
form:
\be \label{e2.1}
\delta \phi_r(x) = \epsilon \, f_r[\{\phi_s\}; x]\, ,
\ee
where $f_r[\{\phi_s\}; x]$ is a functional of the fields $\{\phi_s\}$ and 
a function of $x$, and $\epsilon$ is an infinitesimal parameter. Under 
this transformation $\delta S=0$. 

Now consider a modified transformation law
\be \label{e2.2}
\delta \phi_r(x) = \epsilon(x) \, f_r[\{\phi_s\}; x]\, .
\ee
The only difference between \refb{e2.1} and \refb{e2.2} is that in the 
former $\epsilon$ is a constant whereas in the latter we have taken 
$\epsilon$ to be space-time dependent. In general \refb{e2.2} is not a 
symmetry of $S$. Hence $\delta S$ does not vanish. However it must be 
proportional to $\p_\mu\epsilon$ since it should vanish whenever 
$\epsilon$ is set to a constant. The general expression for $\delta S$ 
will thus have the form
\be \label{e2.3}
\delta S = -\int d^{p+1} x \, \p_\mu\epsilon \, J^\mu (x) = \int d^{p+1} x 
\, 
\epsilon \, \p_\mu\, J^\mu (x)\, ,
\ee
where $J^\mu(x)$ is a function of $x$ and functional of the fields 
$\{\phi_s\}$. Eq.\refb{e2.3} defines the currents $J^\mu(x)$ up to 
addition of a term $K^\mu$ which satisfies $\p_\mu K^\mu=0$ {\it without 
using equations of motion.} This is the 
usual ambiguity in the construction of conserved currents using Noether 
prescription.

So far we have not used equations of motion. If the functions $\phi_r$ 
happen to be solutions of the equation of motion derived from the action 
$S$, then for any variation $\delta\phi_r$ of $\phi_r$ (including the ones 
given in \refb{e2.2}) $\delta S$ must vanish to first order in $\epsilon$. 
Thus in this case we must have
\be \label{e2.4}
\p_\mu J^\mu(x) = 0\, .
\ee
In other words the currents $J^\mu(x)$ defined through \refb{e2.3} are 
conserved when the fields satisfy their equations of motion. This is 
Noether's theorem.

Since we shall be dealing with theories with non-local action,
it is best to work in the momentum space. If we 
define by $\wt\phi_r(k)$, $\wt J^\mu(k)$ and $\wt\eps(k)$ the Fourier 
transforms of $\phi_r(x)$, $J^\mu(x)$ and $\eps(x)$:
\be \label{e2.5}
\wt\phi_r(k) = \int d^{p+1} x\, e^{-i k.x} \, \phi_r(x)\, , \quad
\wt J^\mu(k) = \int d^{p+1} x\, e^{-i k.x} \, J^\mu(x)\, , \quad
\wt \eps(k) = \int d^{p+1} x\, e^{-i k.x} \, \eps(x)\, ,
\ee
then \refb{e2.3} takes the form:
\be \label{e2.6}
\delta S = i \int {d^{p+1} q \over (2\pi)^{p+1}} \, \eps(-q) \, q_\mu \, 
\wt J^\mu(q)\, .
\ee

We shall now use the definition of $J^\mu(x)$ given in \refb{e2.3} to 
construct the energy-momentum tensor $T^\mu_\rho$ of a classical field 
theory with 
non-local action. $T^\mu_\rho$ is the conserved current associated with 
the translational symmetry:
\be \label{e2.7}
\delta \phi_r(x) = \eps \, \p_\rho \, \phi_r(x)\, .
\ee
In order to find $T^\mu_\rho$ we 
consider the modified transformation:
\be \label{e2.8}
\delta\phi_r(x) = \eps(x) \, \p_\rho \, \phi_r(x)\, ,
\ee
or in the momentum space,
\be \label{e2.9}
\delta\wt\phi_r(k) = i\, \int {d^{p+1} q\over (2\pi)^{p+1}} \, \eps(-q) \, 
(k_\rho+q_\rho) \, \wt \phi_r(k+q)\, .
\ee
Then $\wt T^\mu_\rho$ is defined by expressing $\delta S$ under the 
transformation \refb{e2.9} as
\be \label{e2.10}
\delta S = i \, \int {d^{p+1} q\over (2\pi)^{p+1}} \, \eps(-q) \, q_\mu \, 
\wt T^\mu_\rho(q) \, .
\ee
Of course this equation does not necessarily fix $T^\mu_\rho$ uniquely. 
We can add terms of the form $K^\mu_\rho$ to $T^\mu_\rho$ if $\p_\mu 
K^\mu_\rho=0$ {\it without using equations of motion}. Normally this 
ambiguity 
does not affect the definition of the total energy and total momentum if 
the field configuration is regular
so that boundary terms from spatial infinity can be 
ignored during integration by parts.

In order to proceed further, we need to begin with some general form of 
$S$. We express $S$ as a power series expansion in the fields $\phi_r$:
\ben \label{e2.11}
S &=& \sum_{N\ge 2} \, \int {d^{p+1}k_1\over (2\pi)^{p+1}} \, \int{ 
d^{p+1}k_2\over (2\pi)^{p+1}} \, \ldots \int {d^{p+1}k_{N-1}\over 
(2\pi)^{p+1}}\, A^{(N)}_{r_1\cdots r_{N}}(k_1, \ldots k_{N-1}) \nonumber 
\\
&& \qquad 
\phi_{r_1}(k_1) \cdots 
\phi_{r_{N-1}}(k_{N-1}) \, \phi_{r_N}(-k_1-\ldots - k_{N-1})\, .
\een 
Then, under \refb{e2.9}, we have
\ben \label{e2.12}
\delta S &=& i\sum_{N\ge 2} \, \int {d^{p+1} q\over (2\pi)^{p+1}} \, 
\eps(-q) \, \int {d^{p+1}k_1\over (2\pi)^{p+1}} \, \int{
d^{p+1}k_2\over (2\pi)^{p+1}} \, \ldots \int {d^{p+1}k_{N-1}\over
(2\pi)^{p+1}}\, \nonumber \\
&& \bigg[ k_{1\rho}
\, \left\{ A^{(N)}_{r_1\cdots 
r_{N}}(k_1-q, 
\ldots k_{N-1}) - A^{(N)}_{r_1\cdots
r_{N}}(k_1,
\ldots, k_{N-1})\right\} + \cdots \nonumber \\
&& + k_{(N-1)\rho} \, \left\{ 
A^{(N)}_{r_1\cdots
r_{N}}(k_1,
\ldots, k_{N-1}-q) - A^{(N)}_{r_1\cdots
r_{N}}(k_1,
\ldots k_{N-1})\right\} \nonumber \\
&& + q_\rho \, A^{(N)}_{r_1\cdots
r_{N}}(k_1,
\ldots k_{N-1})\bigg]
\, \phi_{r_1}(k_1) \cdots
\phi_{r_{N-1}}(k_{N-1}) \, \phi_{r_N}(-k_1-\ldots - k_{N-1}+q) \nonumber 
\\
&\equiv & i \int {d^{p+1}q\over (2\pi)^{p+1}} \, \eps(-q) \, q_\mu \, 
\wt T^\mu_\rho(q) \, .
\een
This gives an expression for $q_\mu \wt T^\mu_\rho$. From this we can 
extract 
$\wt T^\mu_\rho$ up to the usual ambiguity. A possible prescription, 
which we shall adopt,
is to expand the terms involving $A^{(N)}_{r_1\ldots r_N}$ 
(but not those involving the $\phi_r$'s) in 
Taylor series expansion in $q$, and identify the coefficient of $q_\mu$ 
coming from this expansion in a Lorentz covariant fashion. This gives
\ben \label{e2.possible}
\wt T^\mu_\rho(q) &=& \sum_{N=2}^\infty \, \int \, \prod_{j=1}^{N-1} \, 
{d^{p+1}k_j\over 
(2\pi)^{p+1}} \, \bigg[ \delta^\mu_\rho \, A^{(N)}_{r_1\ldots r_N}
(k_1, \ldots k_{N-1}) \nonumber \\
&& + \sum_{i=1}^{N-1} \, \sum_{s=0}^\infty 
\, k_{i\rho} 
\, (-1)^{s+1} \, {1\over (s+1)!} \, {\p^{s+1} A^{(N)}_{r_1\ldots r_N} 
(k_1, \ldots k_{N-1}) \over \p k_{i\mu} \p k_{i\mu_1} \ldots \p k_{i 
\mu_s} } \, q_{\mu_1} \ldots q_{\mu_s} \bigg] \nonumber \\ && \, 
\phi_{r_1}(k_1) \cdots
\phi_{r_{N-1}}(k_{N-1}) \, \phi_{r_N}(-k_1-\ldots - k_{N-1}+q)\, . 
\nonumber \\
\een
It is easy to verify that $q_\mu \wt T^\mu_\rho$ for $\wt T^\mu_\rho$ 
given in 
\refb{e2.possible} satisfies \refb{e2.12}.

\refb{e2.possible} gives the form of $\wt T^\mu_\rho$ in the general case. 
It is 
instructive to ensure that this gives the correct $T^\mu_\rho$ for a free 
scalar field $\phi$ of mass $m$. In this case only $A^{(2)}(k_1)$ is 
non-zero and has the value ${1\over 2} (-k_1^2 - m^2)$. Substituting this 
into \refb{e2.possible} and Fourier transforming the resulting expression 
we recover the correct $T^\mu_\rho$ for this theory up to the addition of 
a term proportional to $(\p^\mu \p_\rho - \delta^\mu_\rho \square) 
\phi^2$. 
This additional term is conserved without using equations of motion and 
reflects the ambiguity in the definition of $T^\mu_\rho$ using Noether 
prescription. This term does not affect the definition of total energy and 
total
momentum of the system as long as $\phi$ is well behaved at $\infty$.

If we are interested in the total energy-momentum vector 
$P_\rho(x^0)=\int d^p \, x \, T^0_\rho(x)$, 
then their Fourier transform in time -- $\wt P_\rho(q^0)$ -- is given by 
$\wt T^0_\rho(q)|_{\vec q=0}$. \refb{e2.possible} gives:
\ben \label{etotal}
\wt P_\rho(q^0) &=& \sum_{N=2}^\infty \, \prod_{j=1}^{N-1} \, 
{d^{p+1}k_j\over
(2\pi)^{p+1}} \,  
\bigg[ 
\delta^0_\rho \, 
A^{(N)}_{r_1\ldots r_N}
(k_1, \ldots k_{N-1}) \nonumber \\ &&
+ \sum_{i=1}^{N-1} \, {k_{i\rho} 
\over q^0} \left( 
A^{(N)}_{r_1\ldots r_N}(k_1, \ldots k_i-q, \ldots k_{N-1}) - 
A^{(N)}_{r_1\ldots r_N}(k_1, \ldots k_{N-1}) 
\right)\bigg]\nonumber \\ &&
\phi_{r_1}(k_1) \cdots
\phi_{r_{N-1}}(k_{N-1}) \, \phi_{r_N}(-k_1-\ldots - k_{N-1}+q)\Bigg|_{\vec 
q=0} 
\, \, .
\een

We shall now use \refb{e2.possible} to derive the form of $T^\mu_\nu$ for 
the 
special case where the field configuration depends on only one of 
the 
space-time 
coordiates $x$.
We shall take $x$ to be 
a Euclidean coordinate, but the 
case where $x$ is the time coordinate can easily be derived from this by a 
double Wick rotation.\footnote{In this case $T_\rho^\mu$ gives the 
energy-momentum tensor of a spatially homogeneous time dependent 
background.} 
In momentum space we can express $\wt\phi_r(k)$ as: \be \label{e2.13}
\wt\phi_r(k) = (2\pi)^p \, \wt\chi_r(k_x) \, \prod_{\sigma\ne x} 
\delta(k_\sigma) \, 
\, ,
\ee
where $k_x$ is the momentum conjugate to $x$. First consider the case 
$\rho\ne x$ in \refb{e2.possible}. In this case the contribution to the 
terms 
in \refb{e2.possible} involving  
$k_{i\rho}$ vanish due to the delta functions in the $\wt\phi_r(k)$. Thus 
we have 
\ben \label{e2.14}
\wt T^\mu_\rho(q) &=& (2\pi)^p \, \delta^\mu_\rho \, \prod_{\sigma\ne 
x} \, \delta(q_\sigma) 
\, \sum_{N\ge 2} \, \int \prod_{i=1}^{N-1} \, {d k_{ix}\over 2\pi} \,  
A^{(N)}_{r_1\cdots
r_{N}}(k_{1x},
\ldots k_{(N-1)x}) \nonumber \\
&& \, \wt \chi_{r_1}(k_{1x}) \cdots
\wt \chi_{r_{N-1}}(k_{(N-1)x}) \, \wt \chi_{r_N}(-k_{1x}-\ldots - k_{(N-1)x} + 
q_x)\, , \nonumber \\ && \qquad \qquad \qquad \qquad \qquad \qquad 
\qquad \qquad \hbox{for $\rho\ne x$} \, .
\een

Let us also assume that the field configuration has a discrete 
space-time symmetry such that if we 
define
\be \label{e2.15}
A^{(N)}_{r_1\cdots
r_{N}; (i,\sigma)}(k_1,
\ldots k_{N-1}) \equiv {\p\over \p k_{i\sigma}} \, A^{(N)}_{r_1\cdots
r_{N}}(k_1,
\ldots k_{N-1})\, ,
\ee
where $k_{i\sigma}$ denotes the component of the $i$'th 
momentum $k_i$ along the $\sigma$ 
direction,
then for any set $(r_1,\ldots r_N)$ for which 
$\phi_{r_i}$'s do not vanish, we have
\be \label{e2.ex1}
A^{(N)}_{r_1\cdots
r_{N}; (i,\sigma)}(k_1,
\ldots k_{N-1}) = 0\,  \quad \hbox{if $k_{j\sigma}=0$ for every $j$.}
\ee
Note that in evaluating the right hand side of \refb{e2.15} we first 
need to differentiate with respect to $k_{i\sigma}$ for a general set of 
momenta and then set $k_{j\sigma}$ to zero for every $j$ in \refb{e2.ex1}. 
Using \refb{e2.possible}, \refb{e2.13} and \refb{e2.ex1} we can see that 
in this case
\be \label{e2.offd}
T^\mu_x=0 \quad \hbox{for $\mu\ne x$}\, .
\ee

This leaves us to determine 
$T^x_x$. Due to the factors of $\prod_{\sigma\ne x} \delta(k_{\sigma})$ in 
the expression for $\phi_r(k)$ we see that in \refb{e2.possible} we 
shall get a factor proportional to $\prod_{\sigma\ne
x} \, \delta(q_\sigma)$, and hence in this expression for $T^x_x$
all the 
indices $\mu$, $\mu_1,\ldots \mu_s$ can only be equal to $x$. In this case 
we can resum the series expansion to express $\wt T^x_x$ as
\ben \label{e2.resum}
\wt T^x_x(q) &=& (2\pi)^p \, \prod_{\sigma\ne
x} \, \delta(q_\sigma)
\, \sum_{N\ge 2} \, \int \prod_{i=1}^{N-1} \, {d k_{ix}\over 2\pi} \,
\bigg[ A^{(N)}_{r_1\cdots
r_{N}}(k_{1x},
\ldots k_{(N-1)x}) \nonumber \\
&& + \sum_{i=1}^{N-1} \, k_{ix} \, {1\over q_x} \, \left\{ 
A^{(N)}_{r_1\cdots
r_{N}}(k_{1x}, \ldots k_{ix}-q_x, 
\ldots k_{(N-1)x}) - A^{(N)}_{r_1\cdots
r_{N}}(k_{1x},
\ldots k_{(N-1)x}) \right\}\bigg] \nonumber \\ &&
\, \wt \chi_{r_1}(k_{1x}) \cdots
\wt \chi_{r_{N-1}}(k_{(N-1)x}) \, \wt \chi_{r_N}(-k_{1x}-\ldots - k_{(N-1)x} +
q_x)\, . \nonumber \\
\een

The ambiguity in determining $\wt T^\mu_\rho(q)$ using Noether 
prescription corresponds to adding to $\wt T^\mu_\rho(q)$ terms of the 
form $\wt K^\mu_\rho(q) = (q^\mu q_\rho - q^2 \, \delta^\mu_\rho)F(q)$ for 
some $F(q)$. For 
field configurations of the form we are considering $\wt 
K^\mu_\rho(q)$ has support in the subspace of the $q$-space in which 
all components of $q$ other than the $x$-component are set to 
zero. Hence 
$\wt K^x_x=0$, and $\wt T^x_x(q)$ is unambiguously defined in this case.

So far we have not assumed that the field configuration $\wt\chi_r(q_x)$ 
satisfies its equation of motion. However we shall be interested in 
computing $T^\mu_\nu$ only for those configurations which satisfy 
equations of motion. In this case due to the conservation law $\p_\mu 
T^\mu_\rho=0$, and that $T^\mu_x=0$ for $\mu\ne x$, $T^x_x$ must be $x$ 
independent. 
This information can be used to derive various different formul\ae\ for 
$T^x_x$ all of which are physically invariant. The most straightforward 
formula for $T^x_x$ is obtained
by taking the Fourier transform of $\wt T^x_x(q)$:
\ben \label{etnew}
T^x_x(x) &=& \int \prod_\sigma \, {d q_\sigma\over 2\pi}\, e^{i q_\sigma 
x^\sigma} 
\, \wt T^x_x(q)
\nonumber 
\\
&=&  \sum_{N\ge 2} \, \int {d q_x\over 2\pi} \, e^{i q_x 
x} \, \prod_{i=1}^{N-1} \, {d 
k_{ix}\over 2\pi} \,
\bigg[ A^{(N)}_{r_1\cdots
r_{N}}(k_{1x},
\ldots k_{(N-1)x}) \nonumber \\
&& + \sum_{j=1}^{N-1} \, k_{jx} \, {1\over q_x} \, \left\{
A^{(N)}_{r_1\cdots
r_{N}}(k_{1x}, \ldots k_{jx}-q_x,
\ldots k_{(N-1)x}) - A^{(N)}_{r_1\cdots
r_{N}}(k_{1x},
\ldots k_{(N-1)x}) \right\}\bigg] \nonumber \\ &&
\, \wt \chi_{r_1}(k_{1x}) \cdots
\wt \chi_{r_{N-1}}(k_{(N-1)x}) \, \wt \chi_{r_N}(-k_{1x}-\ldots - 
k_{(N-1)x} +
q_x)\, . \nonumber \\
\een
For application to string field theory, a different formula will be 
useful. Since $T^x_x$ is independent of $x$, $\wt T^x_x(q)$ must 
be proportional to $\delta(q_x)$. 
The non-trivial information about $T^x_x$ is then contained in the 
coefficient of the $\delta(q_x)$ in $\wt T^x_x(q)$. In order to extract 
this information we need to evaluate \refb{e2.resum} near $q_x=0$, setting 
$q_x=0$ in all terms which have smooth $q_x\to 0$ limit. Since we shall 
only 
consider theories for which the coefficients $A^{(N)}_{r_1\cdots
r_{N}}(k_{1x},
\ldots k_{(N-1)x})$ are smooth functions of their arguments, we 
get\footnote{Although this expression for $T^x_x$ looks quite different 
from the one
given in \cite{0207107} for $x$ dependent field configurations, the 
expression given in \cite{0207107} can be 
brought to the form given in \refb{e2.newtxx} for on-shell field 
configurations by manipulations involving space-averaging, integration 
by parts and resummation.}
\ben \label{e2.newtxx}
\wt T^x_x(q_x) &=& (2\pi)^p \, \prod_{\sigma\ne
x} \, \delta(q_\sigma)
\, \sum_{N\ge 2} \, \int \prod_{i=1}^{N-1} \, {d k_{ix}\over 2\pi} \,
\bigg[ A^{(N)}_{r_1\cdots
r_{N}}(k_{1x},
\ldots k_{(N-1)x}) \nonumber \\
&& - \sum_{j=1}^{N-1} \, k_{jx} 
\, A^{(N)}_{r_1\cdots
r_{N}; (j,x)}(k_{1x},  
\ldots k_{(N-1)x}) \bigg] \nonumber \\ &&
\, \wt \chi_{r_1}(k_{1x}) \cdots
\wt \chi_{r_{N-1}}(k_{(N-1)x}) \, \wt \chi_{r_N}(-k_{1x}-\ldots - k_{(N-1)x} +
q_x)\, , \nonumber \\
\een
where $ A^{(N)}_{r_1\cdots
r_{N}; (i,x)}(k_{1x},
\ldots k_{(N-1)x})$ have been defined in \refb{e2.15}.
Note that we have not set $q_x=0$ inside the argument of $\wt \chi_{r_N}$ 
in 
anticipation of the fact $\wt\chi_r(k_x)$ may not be a smooth function of 
$k_x$. In fact if 
the $\wt\chi_r(k_x)$ are smooth functions (which is the case if 
$\chi_r(x)$ falls off to zero as $x\to\pm\infty$) then there is no 
possibility of getting a delta function singularity as $q_x\to 0$ in 
\refb{e2.newtxx} and hence $T^x_x$ must vanish. This is expected. Since 
$T^x_x$ must be independent of the coordinate $x$, we can evaluate it at 
$x=\infty$, and hence for a field configuration which vanishes as
$x\to\infty$ 
$T^x_x$ must also vanish.

In order to construct solutions with non-trivial $T^x_x$, we can consider 
periodic field configurations along $x$ direction. If the period is
$2\pi R$ then 
$\chi_r(x)$ must have the form:
\be \label{e2.20}
\chi_r(x) = \sum_{n=-\infty}^\infty \, \wt\chi_{r,n} \, e^{i n x / R} \, .
\ee
This gives
\be \label{e2.21}
\wt\chi_r(q_x) = \int \, dx \, e^{-i q_x x} \, \chi_r(x) = 2\pi \, 
\sum_{n=-\infty}^\infty \, \wt\chi_{r,n} \, \delta\left(q_x-{n\over 
R}\right) 
\, .
\ee
Substituting this into \refb{e2.newtxx} we get near $q_x=0$:\footnote{By 
the general argument leading to the conservation law of $T^\mu_\rho$, the 
contribution to $\wt T^x_x(n/R)$ for $n\ne 0$ vanishes if all the 
$\wt\chi_{r,n}$ 
satisfy their equations of motion.}
\ben \label{e2.22}
\wt T^x_x(q_x) &=& (2\pi)^{p+1} \, \prod_{\sigma} \, 
\delta(q_\sigma)\, \sum_{N\ge 2} \, \sum_{n_1,\ldots n_{N-1}} 
\, \wt\chi_{r_1, n_1} \ldots \wt \chi_{r_{N-1}, n_{N-1}} \, \wt \chi_{r_N, 
-n_1 - \ldots - n_{N-1}} \nonumber \\
&& \bigg[ A^{(N)}_{r_1,\ldots r_N}\left({n_1\over R}, \ldots {n_{N-1}\over 
R}\right) - \sum_{i=1}^{N-1} \, {n_i\over R} \, A^{(N)}_{r_1,\ldots 
r_N; (i,x)}\left({n_1\over R}, \ldots {n_{N-1}\over
R}\right) \bigg] \, . \nonumber \\
\een
After taking Fourier transform this gives
\ben \label{e2.23}
T^x_x &=& \sum_{N\ge 2} \, \sum_{n_1,\ldots n_{N-1}} 
\, \wt\chi_{r_1, n_1} \ldots \wt \chi_{r_{N-1}, n_{N-1}} \, \wt \chi_{r_N, 
-n_1 - \ldots - n_{N-1}} \nonumber \\
&& \bigg[ A^{(N)}_{r_1,\ldots r_N}\left({n_1\over R}, \ldots {n_{N-1}\over 
R}\right) - \sum_{i=1}^{N-1} \, {n_i\over R} \, A^{(N)}_{r_1,\ldots 
r_N; (i,x)}\left({n_1\over R}, \ldots {n_{N-1}\over
R}\right) \bigg] \, . \nonumber \\
\een

For completeness we shall also write down the expression for $T^\mu_\rho$ 
for $\rho\ne x$ in this case. Using \refb{e2.14} and \refb{e2.21} we get
\ben \label{ettri}
\wt T^\mu_\rho(q) &=& (2\pi)^p \, \delta^\mu_\rho \, \prod_{\sigma\ne
x} \, \delta(q_\sigma)
\, \sum_{N\ge 2} \, \sum_{n_1,\ldots n_{N}} \, A^{(N)}_{r_1,\ldots 
r_N}\left({n_1\over R}, \ldots {n_{N-1}\over
R}\right) \nonumber \\
&&
\, \wt\chi_{r_1, n_1} \ldots \wt \chi_{r_{N-1}, n_{N-1}} \, \wt \chi_{r_N,
n_N} \, 2\pi \, \delta\left(q_x- {1\over R} \sum_{i=1}^N \, n_i\right) \, 
, \qquad  \hbox{for $\rho\ne x$}
\, .
\nonumber \\
\een
Taking its Fourier transform gives:
\ben \label{ettri.1}
T^\mu_\rho(x) &=& \delta_\rho^\mu \, \sum_{N\ge 2} \, \sum_{n_1,\ldots 
n_{N}} \, \exp\left( i 
\, {x\over R} \, \sum_{i=1}^{N} n_i\right)
A^{(N)}_{r_1,\ldots
r_N}\left({n_1\over R}, \ldots {n_{N-1}\over
R}\right) \nonumber \\
&&
\, \wt\chi_{r_1, n_1} \ldots \wt \chi_{r_{N-1}, n_{N-1}} \, \wt \chi_{r_N,
n_N} \,
, \qquad  \hbox{for $\rho\ne x$}
\, .
\nonumber \\
\een
The average $T^\mu_\rho$ over a period is given by:
\ben \label{ettri.2}
\la T^\mu_\rho \ra &=& \delta_\rho^\mu \,  \sum_{N\ge 2} \, 
\sum_{n_1,\ldots n_{N-1}} \, 
A^{(N)}_{r_1,\ldots
r_N}\left({n_1\over R}, \ldots {n_{N-1}\over
R}\right)
\, \wt\chi_{r_1, n_1} \ldots \wt \chi_{r_{N-1}, n_{N-1}} \, \wt \chi_{r_N,
-n_1- \ldots -n_{N-1}} \, , \nonumber \\ && \,
\qquad \qquad \qquad \qquad \qquad \qquad \qquad 
\qquad \qquad \qquad 
\hbox{for $\rho\ne x$}
\, .
\een

During the analysis described above,
we 
have given a special role to the $N$'th momentum $k_N$ by replacing it by 
$-(k_1+\ldots k_{N-1})$ and representing $A^{(N)}_{r_1\ldots r_N}$ as a 
function of the momenta $k_1,\ldots k_{N-1}$. For practical computation it 
is often convenient\footnote{For example, if the 
indices 
$r_1,\ldots r_N$ take the same value, then it is more convenient to 
regard $A^{(N)}_{r_1\ldots r_N}$ as a symmetric function of $k_1,\ldots 
k_N$.}
to use all the $N$ arguments $k_1,\ldots k_N$ by
expressing $A^{(N)}_{r_1\ldots r_N}$
as a function 
$\AAA^{(N)}_{r_1,\ldots r_N}$ of $k_1,\ldots k_N$.
(This of course is not unique since we can replace the whole or part of 
the $k_N$ by $-(k_1+\ldots +k_{N-1})$). The 
results in one formalism can be transformed to the other formalism by 
using the equation:
\be \label{e2.24}
{\p\over \p k_{ix}} A^{(N)}_{r_1\cdots r_{N}}(k_{1x}, \ldots k_{(N-1)x})
= {\p\over \p k_{ix}} \AAA^{(N)}_{r_1\cdots r_{N}}(k_{1x}, \ldots k_{Nx})
- {\p\over \p k_{Nx}} \AAA^{(N)}_{r_1\cdots r_{N}}(k_{1x}, \ldots 
k_{Nx})\, 
,
\ee
where it is understood that on both the right and the left hand side we 
take the partial 
derivative keeping all the other arguments fixed. 
This gives: 
\be \label{e2.25}
\sum_{i=1}^{N-1} \, k_{ix} \,  {\p\over \p k_{ix}} A^{(N)}_{r_1\cdots 
r_{N}}(k_{1x}, 
\ldots 
k_{(N-1)x})
= \sum_{i=1}^{N} \, k_{ix} \, {\p\over \p k_{ix}} 
\AAA^{(N)}_{r_1\cdots 
r_{N}}(k_{1x}, \ldots k_{Nx})\, ,
\ee
since $k_{Nx}=-(k_{1x}+\ldots +k_{(N-1)x})$.
Thus if we define
\be \label{e2.26}
\AAA^{(N)}_{r_1\cdots
r_{N}; (i,x)}(k_{1x},
\ldots k_{Nx}) \equiv {\p\over \p k_{ix}} \, \AAA^{(N)}_{r_1\cdots
r_{N}}(k_{1x},
\ldots k_{Nx})\, ,
\ee
we can rewrite \refb{e2.newtxx} and \refb{e2.23} as
\ben \label{e2.27}
\wt T^x_x(q_x) &=& (2\pi)^p \, \prod_{\sigma\ne
x} \, \delta(q_\sigma)
\, \sum_{N\ge 2} \, \int \prod_{i=1}^{N} \, {d k_{ix}\over 2\pi} \,
\quad 2\pi \, \delta (k_{1x}+\ldots k_{Nx}-q_x) \nonumber \\
&& \bigg[ 
\AAA^{(N)}_{r_1\cdots
r_{N}}(k_{1x},
\ldots k_{Nx}) - \sum_{j=1}^{N} \, k_{jx}
\, \AAA^{(N)}_{r_1\cdots
r_{N}; (j,x)}(k_{1x},
\ldots k_{Nx}) \bigg] \nonumber \\ &&
\, \wt \chi_{r_1}(k_{1x}) \cdots
\wt \chi_{r_{N1}}(k_{Nx}) , \nonumber \\
\een
and
\ben \label{e2.28}
T^x_x &=& \sum_{N\ge 2} \, \sum_{n_1,\ldots n_{N}} \, \delta_{n_1+\ldots 
n_N,0}
\, \wt\chi_{r_1, n_1} \ldots \wt \chi_{r_{N}, n_{N}} \nonumber \\
&& \bigg[ \AAA^{(N)}_{r_1,\ldots r_N}\left({n_1\over R}, \ldots 
{n_{N}\over
R}\right) - \sum_{i=1}^{N} \, {n_i\over R} \, \AAA^{(N)}_{r_1,\ldots
r_N; (i,x)}\left({n_1\over R}, \ldots {n_{N}\over
R}\right) \bigg] \, , \nonumber \\
\een
respectively. 
Also, eq.\refb{ettri.2} takes the form
\be \label{ettri.3}
\la T^\mu_\rho \ra = \delta^\mu_\rho \, \sum_{N\ge 2} \, \sum_{n_1,\ldots 
n_{N}} \,
\AAA^{(N)}_{r_1,\ldots
r_N}\left({n_1\over R}, \ldots {n_{N}\over
R}\right)
\, \wt\chi_{r_1, n_1} \, \ldots \, \wt \chi_{r_N,
n_N} \, \delta_{n_1+\ldots n_N, 0} \quad \hbox{for $\rho\ne x$.}
\nonumber \\
\ee
These formul\ae\ look manifestly symmetric in all the 
arguments.

We can further simplify \refb{e2.28}, \refb{ettri.3} as follows. Using 
\refb{e2.11}, 
\refb{e2.13} and \refb{e2.21} we can express the action $S$ as
\be \label{e2.29}
S(\{\chi_{r,n}\}; R) = V_{p+1} \, \sss(\{\chi_{r,n}\}; R) \, ,
\ee
where $V_{p+1} \equiv (2\pi)^{p+1} \delta^{p+1}(k=0)$ denotes the total 
volume of the D-brane world-volume, and
\be \label{e2.29a}
\sss(\{\chi_{r,n}\}; R) = \sum_{N\ge 2} 
\, \sum_{n_1,\ldots n_{N}} 
\AAA^{(N)}_{r_1,\ldots
r_N; (i,x)}\left({n_1\over R}, \ldots {n_{N}\over
R}\right) \, \wt\chi_{r_1, n_1} \ldots 
\wt \chi_{r_{N}, 
n_{N}}\, \, 
\delta_{n_1+\ldots
n_N,0}
\,  ,
\ee 
is the space-time averaged Lagrangian density. Thus we see using 
\refb{ettri.3}
that 
\be \label{ettri.4}
\la T^\mu_\rho\ra = \delta^\mu_\rho \,\sss\, \quad \hbox{for $\rho\ne x$.}
\ee
Also, using \refb{e2.26} we get 
\be \label{e2.30}
\sum_{i=1}^{N} \, {n_i\over R} \, \AAA^{(N)}_{r_1,\ldots
r_N; (i,x)}\left({n_1\over R}, \ldots {n_{N}\over
R}\right) = - R \, {\p\over \p R} \AAA^{(N)}_{r_1,\ldots
r_N}\left({n_1\over R}, \ldots {n_{N}\over
R}\right)\, .
\ee
Using eqs.\refb{e2.29a}, \refb{e2.30} we can express \refb{e2.28} as
\be \label{e2.31}
T^x_x = \sss + R\, {\p \sss\over \p R} = {\p\over \p R} (R\sss)\, ,
\ee
where in \refb{e2.31} the derivative with respect to $R$ has to be 
computed for fixed $\wt\chi_{r,n}$'s. 

In some situations the theory may contain a 
family of solutions, with the periodicity $2\pi R$ of the solution varying 
continuously inside the family. Thus we have a family of solutions
labelled by $R$. In this case we can define $\sss$ to 
be a function of $R$ alone by replacing $\wt\chi_{r,n}$ by their classical 
solutions, and evaluating the expression for $\sss$.
The {\it total derivative} ${d\sss\over dR}$ of $\sss$ will now include 
contribution due to 
explicit $R$ dependence of the coefficients $\AAA^{(N)}_{r_1,\ldots
r_N; (i,x)}\left({n_1\over R}, \ldots {n_{N}\over
R}\right)$ appearing in $\sss$ 
as well as due to the 
$R$ dependence of the solutions
$\wt\chi_{r,n}$. However, since $\{\wt \chi_{s,m}\}$ are 
solutions of the equations of motion, the derivative of $\sss$ with 
respect to $\wt\chi_{r,n}$ vanishes for each $(r,n)$ and ${d \sss\over d 
R}$ is actually equal to ${\p \sss\over \p R}$ appearing in \refb{e2.31}. 
Thus in this case we can 
replace ${\p \over \p R}$ in \refb{e2.31} by total derivative ${d
\over 
d R}$. Since from \refb{ettri.4} we see that $2\pi R \sss$ can be 
interpreted as the integral of $T^0_0$ over a period, if we 
regard $x$ as a compact direction with period $2\pi R$ then $-2\pi 
R\sss$ has the interpretation of the total energy of the solution.
Hence ${d \over dR} (R\sss)$ has the
interpretation of being the negative of the derivative of the total energy
with respect to the period. This precisely coincides with the usual 
definition of pressure 
of any system.

\sectiono{Application to $p$-adic String Theory} \label{spadic}

In this section we shall apply the results of section \ref{senergy}
to derive the form of the stress tensor in $l$-adic string theory and 
compare with the results of \cite{0207107,0209197}. We begin with the 
action of the $l$-adic string theory\cite{padic,FROK,0003278}:
\be \label{ep.1}
S = \int d^{p+1} x \, \LL, \qquad \LL = {1\over g_l^2} \, 
\left[-{1\over 2} \, \phi \, l^{-{1\over 2} \Box} \, \phi + {1\over l+1} 
\, \phi^{l+1}\right]\, ,
\ee
where $l$ is an integer, $\phi$ is a scalar field and $g_l$ is the 
coupling constant of the open 
$l$-adic string theory.
Comparing this with eq.\refb{e2.11} we see that in this case:
\ben \label{ep.2}
&& A^{(2)}(k_1) = -{1\over 2 g_l^2} \,  l^{{1\over 2} k_1^2}, \qquad
A^{(l+1)}(k_1, \ldots k_l) =  {1\over g_l^2} \, {1\over l+1} \, , 
\nonumber \\
&& A^{(N)}_{r_1\ldots r_N}(k_1,\ldots k_{N-1}) =0 \quad \hbox{for 
$N\ne 2, 
l+1$}\, . 
\een
For the case of odd $l$, 
ref.\cite{0207107} constructed a one parameter family of time dependent 
solution of the form:
\be \label{ep.3}
\phi(x^0) = \sum_{n=0}^\infty \, a_{2n+1} \, \cos\left( (2n+1) \omega 
x^0\right)\, ,
\ee
labelled by the arbitrary constant $\omega$. For any given $\omega$ the 
coefficients $a_{2n+1}$ are determined from the equations of motion. We 
shall now find an 
expression for $T_0^0$ for such a field configuration using the general 
formalism given in section \ref{senergy} and compare with the 
corresponding expression obtained in \cite{0207107}.

We derived various formul\ae\ for $T_x^x$ for an $x$-dependent field 
configuration in section \ref{senergy}. We can take any of these and make 
an inverse Wick rotation $x\to - i x^0$ to get the answer for $T^0_0$ for 
a time dependent field configuration. The formula that directly yields 
the result of \cite{0207107} is eq.\refb{etnew} at $x=0$, giving the 
expresstion for 
$T^0_0$ at $x^0=0$ after inverse Wick rotation. The first term inside the 
square bracket on the right 
hand side of \refb{etnew} just gives the lagrangian density $\LL$ 
evaluated at 
$x^0=0$. Since the other terms involve differences in $A^{(N)}$ evaluated 
for two different momentum arguments, and since from \refb{ep.2} we see 
that $A^{(l+1)}$ does not depend on momenta, this term receives 
contribution only for $N=2$. Using \refb{ep.1}-\refb{ep.3},
\refb{etnew} takes the form:
\be \label{ep.4}
T^0_0 = \LL(x^0=0) \, - {1\over 2 g_l^2} \, \sum_{m,n\ge 0} \, a_{2m+1} \, 
a_{2n+1} \, 
{(2m+1)^2 \over (2m+1)^2 - (2n+1)^2 } \, \left( l^{-{1\over 2} (2n+1)^2 
\omega^2 } - l^{-{1\over 2} (2m+1)^2
\omega^2 } \right)\, .
\ee
Note that the $m=n$ term in the above expression has 0/0 form and hence 
must be defined as the result of taking $m\to n$ limit. This gives a 
modified expression for \refb{ep.4}
\ben \label{ep.5}
T^0_0 &=& \LL(x^0=0) \, - \, {1\over 4 g_l^2} \, \omega^2 
\, \ln l \, \sum_{n=0}^\infty \, 
a_{2n+1}^2 \, (2n+1)^2 \, l^{-{1\over 2} (2n+1)^2
\omega^2 }
\nonumber \\
&& - {1\over 2 g_l^2} \, \sum_{m,n\ge 0\atop m\ne n} \, a_{2m+1} \,
a_{2n+1} \,
{(2m+1)^2 \over (2m+1)^2 - (2n+1)^2 } \, \left( l^{-{1\over 2} (2n+1)^2
\omega^2 } - l^{-{1\over 2} (2m+1)^2
\omega^2 } \right)\, . \nonumber \\
\een
This agrees with the result derived in \cite{0207107}. One simplicity 
of our starting formula \refb{etnew} is that we arrive at the answer 
\refb{ep.5} 
without 
having to do any resummation as in \cite{0207107}. Using the equations 
of motion $\phi^l = l^{-{1\over 2}\Box}\phi$, and \refb{ep.1}, \refb{ep.3} 
we can write down a simple expression for $\LL(x^0=0)$ appearing in 
\refb{ep.5}:
\be \label{ep.f}
\LL(x^0=0) = {1\over 2 g_l^2} \, {1-l\over 1+l} \, \sum_{m,n\ge 0} 
a_{2m+1} \, a_{2n+1} \, l^{-{1\over 2} \omega^2 \, (2n+1)^2}\, .
\ee

\sectiono{Application to Lump Solutions in String Field Theory} 
\label{sappli}

In this section we shall apply the results of section \ref{senergy} to
calculate $T^x_x$ associated with the lump solutions in open string field
theory \cite{0005036,0101014}. As in \cite{0005036}, we shall
focus on a D-string of bosonic string
theory with the direction $x$ tangential to the D-string compactified on a 
circle of radius $R$. The action of the open string field theory on the 
D-string is given by:
\be \label{e3.1}
S = -{1\over g_o^2} \, \left[ {1\over 2} \la \Phi| Q_B |\Phi\ra + {1\over 
3} \la\Phi| \Phi * \Phi\ra \right],
\ee
where $|\Phi\ra$ is the string field represented by a ghost number 1 state 
in the Hilbert space of the first quantized string theory, $*$ denotes the 
usual open string star product\cite{WITTENSFT} and $g_o$ is the open 
string coupling constant.
In \cite{0005036} the space averaged lagrangian density
associated with a time independent, periodic field configuration 
was denoted by:
\be \label{e3.1a}
\sss(|\Phi\ra; R) = -2\pi^2 \, \TT_1 \VV(|\Phi\ra; R)\, .
\ee
Unlike in the case of ref.\cite{0005036} here we have explicitly 
displayed the $R$ dependence of $\VV$. 
Let us denote by 
$|\Phi_{lump}\ra$ the lump solution in this string field theory, 
representing a D0-brane 
localized at $x=0$.
Using 
\refb{e2.31} we can express $T^x_x$ associated with this solution as:
\be \label{e3.1b}
T^x_x(|\Phi_{lump}\ra) = -2\pi^2 \, \TT_1 \, \left(\VV(|\Phi_{lump}\ra; R) 
+ R \, \VV'(|\Phi_{lump}\ra; R) \right)\, ,
\ee
where
\be \label{edefvvp}
\VV'(|\Phi\ra; R) \equiv {\p 
\VV(|\Phi\ra; R) \over \p R} \, ,
\ee
with $\p/\p R$ denoting derivative with respect to $R$ at fixed 
$|\Phi\ra$.
Since the 
original D-string has $T^x_x=-\TT_1$, and since the final D0-brane should 
have vanishing $T^x_x$, we see that the $T^x_x(|\Phi_{lump}\ra)$ computed 
from the string 
field theory must cancel the $T^x_x$ of the D-string. This leads to the 
conjecture:
\be \label{e3.4a}
T^x_x(|\Phi_{lump}\ra) = \TT_1 \, .
\ee
Using \refb{e3.1b} this takes the form:
\be \label{e3.4b}
-2\pi^2 \, \left(\VV(|\Phi_{lump}\ra; R)
+ R\, \VV'(|\Phi_{lump}\ra; R) \right) =1\, .
\ee

On the other hand from \refb{ettri.4}
we see that the 
$T^0_0$ associated with the solution is equal to the average lagrangian 
density $-2\pi^2 \TT_1 \VV(|\Phi_{lump}\ra;R)$. Thus 
the total energy associated with the lump solution, obtained by 
multiplying 
$-T^0_0$ by $2\pi R$, and adding to it the energy $2\pi R\, \TT_1$ of the 
original 
D-string, will be given by\cite{0005036}:
\be \label{e3.4c}
2 \pi R \, \TT_1 \, \left(2\pi^2 \, \VV(|\Phi_{lump}\ra; R) + 1\right) 
\, .
\ee
If $|\Phi_{lump}\ra$ has to describe a D0-brane then its energy 
must be equal to the expected mass of the D0-brane which
is $2\pi \TT_1$. Thus the energy conjecture 
implies\cite{0005036}
\be \label{e3.4d}
R\left(2\pi^2 \, \VV(|\Phi_{lump}\ra; R) + 1\right) = 1\, .
\ee

\refb{e3.4d} was verified numerically in \cite{0005036}, and using the 
expression of $\VV(|\Phi\ra; R)$ and the numerical solution for 
$|\Phi_{lump}\ra$ given there we can also verify 
\refb{e3.4b} numerically. But
we shall now show that the pressure conjecture \refb{e3.4b} follows 
automatically from the energy conjecture \refb{e3.4d}.
For this we differentiate \refb{e3.4d} with respect to $R$ to 
get
\be \label{e3.4e}
2\pi^2 \, \left(\VV(|\Phi_{lump}\ra; R)
+ R\, {d
\VV(|\Phi_{lump}\ra; R) \over d R} \right) +1=0\, ,
\ee
where the derivative with respect to $R$ now also acts on the various $R$ 
dependent coefficients appearing in the expansion of $|\Phi_{lump}\ra$.
However, using the argument at the end of section \ref{senergy} we can 
replace 
$  {d
\VV(|\Phi_{lump}\ra; R) / d R}$ by
$ {\p
\VV(|\Phi_{lump}\ra; R) / \p R} = \VV'(|\Phi_{lump}\ra; R)$ where the $R$ 
derivative acts only on the coefficients appearing in 
$\VV((|\Phi\ra; R)$ at fixed $|\Phi\ra$. Eq.\refb{e3.4e} then reduces to 
\refb{e3.4b}.

Thus we see that for the tachyon lump solution in open string field 
theory, the vanishing of the total pressure in direction transverse to the 
lump 
is not an independent conjecture, but follows from the conjecture that the 
total energy of the lump solution is given by the mass of the D0-brane, 
which in 
turn is independent of the radius of the circle. While this does not lead 
to a new test of the validity of string field theory, this shows that the 
formula for $T^x_x$ for a periodic solution, as given in \refb{e2.31}, is 
correct. In section \ref{smarginal} we shall use \refb{e2.31} to relate 
the marginal deformation parameter in the conformal field theory 
description of a solution to the 
parameter labelling the corresponding solution in string field theory.

\sectiono{Marginal Deformation in Open String Field Theory} 
\label{smarginal}

In section \ref{sappli} we calculated $T^x_x$ associated with the lump 
solution in open string field theory for a D-string on a circle of radius 
$R$ and showed that 
the result agrees with the prediction based on the identification of the 
solution as a D0-brane. In this section we shall consider the same system 
at $R=1$. In this case instead of having a unique lump solution there is a 
one parameter family of lump solutions. In the language of conformal field 
theory this corresponds to adding to the world-sheet action a boundary 
term:
\be \label{e4.1}
\tl \, \int dt \, \cos\left(  X(t)\right)\, ,
\ee
where $X$ is the world-sheet scalar field describing the compact 
coordinate, $\tl$ is a parameter labelling the deformed conformal field 
theory and $t$ is a 
parameter 
labelling the boundary of the world-sheet. Total $T^x_x$ associated with 
this 
solution can be calculated from the boundary state
associated with 
the 
deformed conformal field theory\cite{9402113,9811237} and takes the 
value\cite{0203211,0203265}
\be \label{e4.2}
T^x_x|_{CFT} = -\TT_1 \, \cos^2 (\pi\tl)\, .
\ee

Since we expect a one to one correspondence between the classical 
solutions of open 
string field equations and boundary conformal field theories, we expect 
that the open string field theory on a D-string wrapped on a circle of 
unit radius must also have a one parameter family of classical solutions. 
A systematic procedure for constructing these classical solutions using 
level truncation approximation\cite{KS,KP,9912249,0002237} was developed 
in 
\cite{0007153}. We 
look for a string field configuration of the form:
\be \label{e4.3}
|\Phi\ra = \sum_{n=0}^\infty \, \left[ t_n \, c_1 + u_n \, c_{-1} + v_n \, 
c_1 \, L^X_{-2} + w_n \, L'_{-2} + \cdots \right] \, \cos\left({n\over R} 
X(0)\right) 
|0\ra\, ,
\ee
subject to the Siegel gauge condition, periodicity along $x$, 
invariance under twist and $x\to -x$ transformation etc. Here $t_n$, 
$u_n$, 
$v_n$ etc. are coefficients labelling a given string field configuration, 
$c_n$, $b_n$ are oscillators of the ghost fields, $L^X_{n}$ denote the 
Virasoro generators of the $c=1$ conformal field theory associated with 
the $X$ field, $L'_n$ denote the Virasoro generators of the $c=25$ 
conformal field theory involving rest of the fields, and $\cdots$ 
stand for infinite number of other terms involving other ghost number one 
operators constructed out of $c_n$, $b_n$, $L^X_n$ and $L'_n$. 

A given 
classical solution in this theory corresponds to a fixed set of values for 
the coefficients $t_n$, $u_n$, $v_n$ etc. Thus a one parameter family of 
solutions will correspond to a one parameter family of 
$\{t_n,u_n,v_n,\ldots\}$. To leading order 
in $\tl$ the solution corresponds to
\be \label{e4.4}
t_1 \simeq \tl\, ,
\ee
with all other coefficients being zero. This suggested the following 
procedure for constructing the one parameter family of solutions. For a 
generic $R$, we first 
fix a value of $t_1$, and solve for the other coefficients by using their 
equations of motion. Plugging these solutions back into the action we get 
an effective action $-2\, \pi^2\, \VV_{eff}(t_1;R)$ for the coefficient 
$t_1$. 
For general $R>1$, $\VV_{eff}(t_1;R)$ will have a local minimum at some 
value of 
$t_1$, and the lump solution will correspond to this minimum. However
if there 
really exists a 
one parameter familty of solutions for $R=1$ then $\VV_{eff}(t_1;1)$
should be 
independent of $t_1$, and hence must vanish since it vanishes at $t_1=0$. 
In this case the one parameter family of solutions will be labelled by 
$t_1$. It was found using the level truncation analysis 
that while to any finite level of approximation $\VV_{eff}(t_1;1)$ is 
not flat,
it does become flatter as we increase the level of approximation. This 
suggested that the full open string field theory does have a one parameter 
family of classical 
solutions representing the conformal field theories associated with 
\refb{e4.1}.

One of the problems which was left unresolved in \cite{0007153} was to
develop a systematic scheme for relating the parameter $t_1$ labelling 
the solution in open string field theory to the conformal field 
theory parameter $\tl$ appearing in eq.\refb{e4.1} beyond the leading 
order result \refb{e4.4}. 
This is the problem we address in this section.\footnote{An analytic 
expression for the solution was proposed in \cite{0107046}. This 
expression is a sum of a term linear in $\tl$, and a term quadratic in 
$\tl$. Since the open string field theory action is a cubic polynomial in 
the string field, the energy momentum tensor calculated for this 
configuration will be a polynomial function of $\lambda$ of degree 6. It 
is not clear how this can reproduce the $\cos^2(\pi\tl)$ factor in the 
expression for $T^x_x$ given in \refb{e4.2}.} This will be done 
by comparing the expression for $T^x_x$ given in \refb{e4.2} with the 
$T^x_x$ associated with the lump solution given in \refb{e3.1b} plus the 
contribution $-\TT_1$ of the original D-string. This gives:
\be \label{e4.6}
-\cos^2 (\pi\tl) = -1 -2\pi^2 \, \left(\VV_{eff}(t_1;1) + 
\VV'_{eff}(t_1;1)\right)\, ,
\ee
or, equivalently,
\be \label{e4.7} 
\sin^2 (\pi\tl) = -2\pi^2 \, \left(\VV_{eff}(t_1; 1)
+ \VV'_{eff}(t_1; 1)\right) \equiv F(t_1)\, ,
\ee
where 
$\VV'_{eff}(t_1;R)$ denotes the result of replacing in 
$\VV'(|\Phi\ra, R)$ all coefficients 
other than $t_1$ by solution to their equations of motion. This is 
equivalent to defining $\VV'_{eff}(t_1;R)$ as
\be \label{edefvvnew}
\VV'_{eff}(t_1;R) = {\p\over \p R} \VV_{eff}(t_1;R)\, ,
\ee
with the derivative computed at fixed $t_1$.
Apparently the $\p/\p R$ on the right hand side of eq.\refb{edefvvnew} 
receives additional contribution besides $\VV'(|\Phi\ra, R)$ due to 
the implicit $R$-dependence of $\VV_{eff}(t_1;R)$ through the various 
coefficients 
other than $t_1$ which were eliminated by their equations of motion.
However since in the construction of $\VV_{eff}(t_1;R)$, $\VV(|\Phi\ra; 
R)$ is extremized with respect to these 
coefficients, these additional terms do not contribute, and 
eq.\refb{edefvvnew} holds.

For real $\lambda$, the left hand side of eq.\refb{e4.7} cannot exceed 
unity. Thus for consistency the right hand side, which is a specific 
expression in string field theory, must also be bounded from above by 
unity. We shall now argue that this is indeed the case.
In particular, assuming that the energy conjecture 
\refb{e3.4d} holds, and that $\VV_{eff}(t_1;1)$ is independent of $t_1$ 
and hence vanishes, we shall give an 
analytical proof of the 
fact that the function $F(t_1)$ defined in eq.\refb{e4.7} has a 
maximum 
where it takes the value 1, and hence $\tl$ defined through 
\refb{e4.7} is real. For this let us define by $t_1^{(0)}(R)$ 
the value of $t_1$ that minimizes $\VV_{eff}(t_1,R)$ with respect to $t_1$ 
for a generic $R$. 
This definition does not work for $R=1$ since there $\VV_{eff}(t_1;1)$ is 
flat, but we shall define by $t_1^{(0)}(1)$ the limit of 
$t_1^{(0)}(R)$ as $R\to 1$. We shall now show that
\begin{enumerate}
\item $F(t_1)$ reaches a maximum at $t_1=t_1^{(0)}(1)$.
\item $F(t_1^{(0)}(1))=1$.
\end{enumerate}
This will establish that the maximum value $F(t_1)$ can take is 1, and 
hence $\sin^2(\pi\tl)$ defined through \refb{e4.7} gives real $\lambda$ 
for all $t_1$.
This will also show that the point $\lambda=1/2$ 
corresponds to $t_1=t_1^{(0)}(1)$, {\it i.e.} the configuration in string 
field theory representing a D0-brane, since for a generic $R$ the lump 
solution represents a D0-brane. This is the expected 
result.\footnote{Before getting into the formal proof it is worth giving 
an intuitive argument. Since $\VV_{eff}(t_1;1)$ is expected to vanish, we 
have $-2\pi^2\, \VV_{eff}(t_1;1+\epsilon) = -2 \pi^2 \, \epsilon \, 
\VV'_{eff}(t_1;1) + \OO(\epsilon^2) = \epsilon \, F(t_1) + 
\OO(\epsilon^2)$. Thus the maximum of $F(t_1)$ occurs at the minimum of 
$\VV_{eff}(t_1;1+\epsilon)$ for small $\epsilon$, {\it i.e.} at 
$t_1^{(0)}(1)$. The energy conjecture \refb{e3.4d} tells us that $2\pi^2\, 
\VV_{eff}(t_1;1+\epsilon) = (1+\epsilon)^{-1}-1=-\epsilon + 
\OO(\epsilon^2)$ at its minimum $t_1^{(0)}(1+\epsilon)\simeq 
t_1^{(0)}(1)$. This gives 
$F(t_1^{(0)}(1))=1$. \label{fo1}}

We begin by proving the first result. Since $\VV_{eff}(t_1;1)$ vanishes, 
we get from \refb{e4.7}, \refb{edefvvnew}
\be \label{e6.1}
{d F(t_1)\over d t_1} = - 2\, \pi^2 \, {\p \VV'_{eff}(t_1, 1)\over 
\partial t_1}
= - 2 \, \pi^2 \, \left[{\p\over \p R} \, \left({\p \VV_{eff}(t_1; R)\over 
\p
t_1}\right)\right]_{R=1} \, .
\ee
Substituting $t_1=t_1^{(0)}(1)$ on both sides we get:
\ben \label{e6.2}
&& {d F(t_1)\over d t_1}\bigg|_{t_1=t_1^{(0)}(1)} = - 2 \, \pi^2 \, 
\left[{\p\over \p R} \, \left({\p \VV_{eff}(t_1; R)\over
\p
t_1}\bigg|_{t_1=t_1^{(0)}(1)}\right)\right]_{R=1} \nonumber \\
&& \qquad = - 2 \, \pi^2 \,
\left[{d\over d R} \, \left({\p \VV_{eff}(t_1; R)\over
\p
t_1}\bigg|_{t_1=t_1^{(0)}(R)}\right)- \left({\p^2 
\VV_{eff}(t_1; R)\over \p 
t_1^2}\right)_{t_1=t_1^{(0)}(R)} \, {d 
t_1^{(0)}(R)\over dR} \right]_{R=1} \, .\nonumber \\
\een
Now, by definition $(\p \VV_{eff}(t_1; R) /
\p
t_1)|_{t_1=t_1^{(0)}(R)}$ vanishes for all $R$. On the other hand, at 
$R=1$, $\VV_{eff}(t_1;R)$ vanishes identically, and hence its second 
derivative with respect to $t_1$ also vanishes. Thus we see that the right 
hand side of \refb{e6.2} vanishes identically, and hence $F(t_1)$ has a 
extremum at $t_1=t_1^{(0)}(1)$. The fact that it is a maximum and 
not a minimum can be seen easily by differentiating \refb{e6.1} once more 
with respect to $t_1$ and using the fact that for $R>1$, 
$\VV_{eff}(t_1;R)$ has a minimum at $t_1=t_1^{(0)}(R)$.

It now remains to prove that $F(t_1^{(0)}(1))=1$. 
Using eq.\refb{e4.7} and the fact that $\VV_{eff}(t_1,1)$ vanishes 
identically we see that
\be \label{e6.3a}
F(t_1^{(0)}(1)) = - 2\, \pi^2 \, \VV'_{eff}(t_1^{(0)}(1), 1)\, .
\ee
Using the energy 
conjecture \refb{e3.4d}
and that fact that 
$t_1=t_1^{(0)}(R)$ describes the lump solution at radius $R$, we have:
\be \label{e6.4}
R \left( 2\pi^2 \, \VV_{eff}(t_1^{(0)}(R); R) + 1 \right) = 1\, .
\ee
Differentiating this with respect to $R$, and noting that $R$ derivative 
action on $t_1^{(0)}(R)$ can be neglected since $\VV_{eff}(t_1,R)$ is 
extremized 
with respect to $t_1$ at $t_1=t_1^{(0)}(R)$, we get
\be \label{e6.5}
2\pi^2 \, \VV_{eff}(t_1^{(0)}(R); R) + 1 + 2\pi^2 \, R \, 
\VV'_{eff}(t_1^{(0)}(R); R) = 
0\, .
\ee
Setting $R=1$ in this equation makes the first term vanish identically. 
Thus the equation reduces to
\be \label{e6.6}
2\pi^2 \, \VV'_{eff}(t_1^{(0)}(1); 1) = -1\, .
\ee
Hence from \refb{e6.3a}, \refb{e6.6} we have
\be \label{e6.7}
F(t_1^{(0)}(1)) = 1\, .
\ee
This is the desired result.

We shall now describe the explicit computation of the function $F(t_1)$ 
using level 
truncation 
approximation. Before getting into the numerical results it will be 
instructive 
to illustrate the computation of $F(t_1)$ using \refb{e4.7} in a 
simple approximation where analytic solution is 
possible. This is level $(1/R^2, 2/R^2)$ approximation.\footnote{Note that 
although eventually we shall set $R=1$, we must keep $R$ arbitrary, 
calculate $\p\VV/\p R$, and only then set $R=1$.} In this approximation 
the only modes to be included are $t_0$ and $t_1$, and $\VV(|\Phi\ra;R)$ 
is given by:
\be \label{e4.8}
\VV(|\Phi\ra;R) = -{1\over 2} t_0^2 - {1\over 4} \, \left( 1 - {1\over 
R^2}\right) \, 
t_1^2 + {1\over 3} \, K^3 \, t_0^3 + {1\over 2} \, K^{3 - {2\over R^2}} \, 
t_0 t_1^2\, ,
\ee
where
\be \label{edefk}
K = {3\sqrt{3}\over 4}\, .
\ee
This gives
\be \label{e4.8a}
R \, \VV'(|\Phi\ra;R) = -{1\over 2} \, {1\over
R^2} \,
t_1^2 + {1\over 2} \, {4\over R^2} \, \ln K \, K^{3 - {2\over R^2}} \,
t_0 t_1^2\, .
\ee
The equation $\p \VV/ \p t_0=0$ has two solutions for $t_0$:
\be \label{e4.9}
t_0 = {1\over 2 K^3} \, \left( 1 \pm \sqrt{1 - 2 K^{6 - {2\over R^2}} 
t_1^2}\right)\, .
\ee
Of these the solution with the + sign has the property that $t_0$ does not 
vanish even when $t_1=0$. This branch was called the vacuum branch in 
\cite{0007153} (since for $t_1=0$ it describes the tachyon vacuum 
solution) and is not the branch of interest to us. The other branch, 
called the marginal branch, gives 
the solution we are looking for. Thus we take
\be \label{e4.10}
t_0 = {1\over 2 K^3} \, \left( 1 - \sqrt{1 - 2 K^{6 - {2\over R^2}}
t_1^2}\right)\, .
\ee
For this solution $t_0$ vanishes at $t_1=0$. Substituting this back into 
\refb{e4.8} and \refb{e4.8a} we get $\VV_{eff}(t_1;R)$ and 
$R\,\VV'_{eff}(t_1;R)$ respectively.
This $\VV_{eff}(t_1;R)$ of course 
is not 
flat, but for $R=1$ it approaches a flat potential as we increase the 
level of 
approximation\cite{0007153}. Note also that real solution for $t_0$ exists 
only for 
$|t_1|\le 1/\sqrt 2 K^{3 - R^{-2}}$. At $|t_1|= 1/\sqrt 2 K^{3 - 
R^{-2}}$ the vacuum branch 
and the marginal branch 
meet. This feature survives even at higher level of approximation 
so that we always get a finite range of $t_1$ over which the solution 
exists\cite{0007153}.

\begin{figure}[!ht]
\leavevmode
\begin{center}
\epsfysize=5cm
\epsfbox{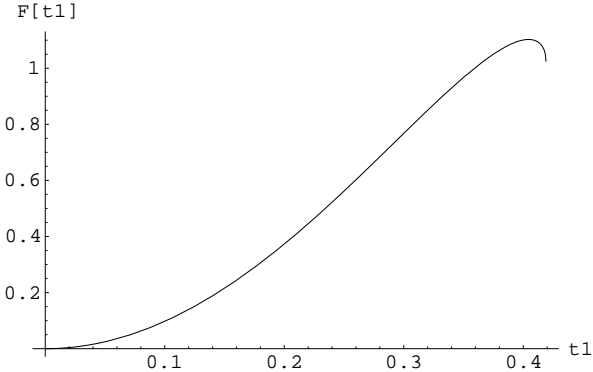}
\end{center}
\caption{The plot of $\sin^2(\pi\tl)=F(t_1)$ vs. $t_1$ at level (1,2) 
calculated 
using 
\refb{e4.11}.
} \label{f1}
\end{figure}
We now use eq.\refb{e4.7}, \refb{e4.8} and \refb{e4.8a} to find $F(t_1)$ 
and hence the 
relation between $\tl$ and $t_1$. 
We get
\ben \label{e4.11}
F(t_1) &=& -2\pi^2 \, \left[ -{1\over 2} t_0^2 - {1\over 4} \, 
\left( 1 + {1\over R^2}\right) \,
t_1^2 + {1\over 3} \, K^3 \, t_0^3 + {1\over 2} \, \left(1 + {4\over 
R^2} \, \ln K\right) \, K^{3 - 
{2\over R^2}} \,
t_0 t_1^2
\right]_{R=1} \nonumber \\
&=& 2\pi^2 \, \left[ {1\over 2} t_0^2 + {1\over 2} \,
t_1^2 - {1\over 3} \, K^3 \, t_0^3 - {1\over 2} \, (1 + 4\ln K) \, K
\, t_0 t_1^2
\right]
\een
Substituting the expression for $t_0$ from \refb{e4.10} at $R=1$ we get 
$F(t_1)$ as 
a 
function of $t_1$. In Fig.\ref{f1} we have displayed a plot of 
$F(t_1)$ vs. $t_1$ calculated using this formula. From this we see 
that the right hand side of \refb{e4.11} becomes larger than one
even before 
$t_1$ reaches the critical point where the marginal and the vacuum 
branches meet, and hence $\tl$ calculated from \refb{e4.7} becomes 
complex in this region. 
Since we argued earlier that $F(t_1)$ is bounded from above by 1, we 
expect this to be
an artifact of 
the level truncation approximation.

\begin{figure}[!ht]
\leavevmode
\begin{center}
\epsfysize=5cm
\epsfbox{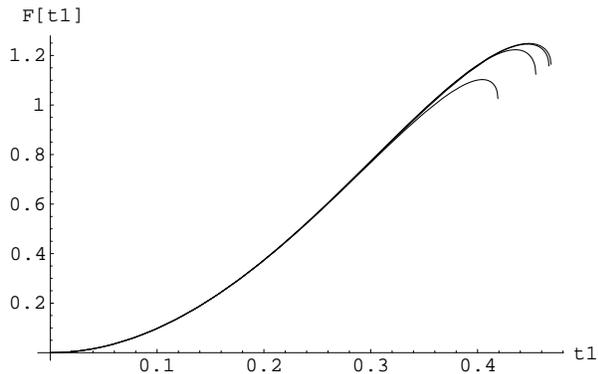}
\end{center}
\caption{The plot of $\sin^2(\pi\tl)=F(t_1)$ vs. $t_1$ at level (1,2), 
(2,4), 
(3,6) and (4,8) approximation. The graph which extends further to the 
right corresponds to result in a higher level approximation. The level 
(3,6) and (4,8) results are almost indistinguishible in this graph. We see 
that the result exceeds unity beyond a certain value of $t_1$, and hence 
beyond this value the $\tl$ defined through \refb{e4.7} becomes complex.} 
\label{f3}
\end{figure}
Calculation at higher level can be carried out using numerical methods. 
The full $R$ dependent tachyon potential needed for calculations up to 
level (4,8) was given in \cite{0007153} (appendix B). With the help of 
this result and eq.\refb{e4.7} we can easily compute $F(t_1)$ as a 
function of $t_1$ at various levels of approximation up to level (4,8). 
The results have been shown in 
Fig.\ref{f3}. In this diagram 
the graph which extends further to the
right corresponds to result in a higher level approximation. 
We see from this graph that
the result for $F(t_1)$ again exceeds unity beyond a certain value of 
$t_1$, and hence
beyond this value the $\tl$ defined through \refb{e4.7} becomes complex. 
This seems to be in conflict with the general result proved earlier that 
$F(t_1)$ is bounded from above by unity. We believe this is due to 
the fact that we have not reached a sufficiently high level of 
approximation. At least the results show that the growth of the maximum 
(which at level (4,8) is slightly above 1.2) slows down with increasing 
level of approximation. Thus it could turn around at higher level. Clearly 
we need explicit numerical results at higher level to settle this issue. 
The
extrapolation metods of refs.\cite{0208149,0211012} may also be useful for 
this 
study.

Although we have not resolved the numerical problem, we have tried to
isolate the origin of the problem. From \refb{e4.7} we see that there are
two possible sources of error. The $\VV_{eff}(t_1;1)$ which
is supposed to vanish identically is not zero at finite level, and the
$2\pi^2\VV'_{eff}(t_1;1)$ which is supposed to take the value $-1$ at its
minimum, is not actually $-1$ at a finite level. It turns out that the
contribution from the first term is negligible, and almost all the error 
comes
from the second term. In particular, if we calculate
$2\pi^2(\VV_{eff}(t_1;R) - \VV_{eff}(t_1;1)) / (R^{-1} -1)$, then
according to the energy conjecture \refb{e3.4d} and the vanishing of
$\VV_{eff}(t_1;1)$ this is supposed to take the value 1 at its maximum.  
Instead we find that for values of $R$ close to unity, this ratio is about
$1.2$ at its maximum in level (4,8)  approximation. If we believe 
that the energy conjecture \refb{e3.4d} is satisfied, then this ratio must 
come down to unity at higher level approximation, and consequently the 
maximum value of 
$F(t_1)$ will also come down to unity.

One feature that we observe from Fig.\ref{f3} is that the maximum of 
$F(t_1)$ is quite close to the critical value of $t_1$ where the level 
truncation method breaks down.
Assuming that the maximum does come down 
to unity at higher level in agreement with the analytical result, it is 
tempting to speculate that the critical value of $t_1$ coincides with this 
maximum and hence describes the $\tl={1\over 2}$ point. 
Since the range $-{1\over 2} \le \tl\le {1\over 2}$ can be thought of as 
the fundamental domain in the moduli space of lump solutions (with 
solutions outside this range being equivalent to the solutions in this 
range) this would imply that string field theory lump solutions precisely 
cover one fundamental domain. Only more detailed 
numerical analysis can tell us if this is so. However it is amusing to 
note that for superstring theory this is precisely what 
happens\cite{0312003} for 
the 
periodic solutions 
of the 
effective action given in \cite{0303139}.

\medskip

{\bf Acknowledgement}: I wish to thank B.~Zwiebach for useful discussions 
and many helpful comments on the manuscript.

\end{document}